%% file: CVComputingWQuantumMemoriesArxiv.tex
\documentclass[aps,pra,reprint,floatfix,superscriptaddress]{revtex4-1}
\usepackage{amsmath,amssymb}    
\usepackage{graphicx}   
\usepackage[dvipdfx,cmyk]{xcolor}     
\usepackage[caption=false]{subfig}  
\usepackage[colorlinks,linkcolor=red,citecolor=brown,urlcolor=blue]{hyperref}
\usepackage{placeins}
\usepackage{dsfont,bm}
\usepackage{color}

\input{header.tex}

\newcommand{\be}{\begin{equation}}
\newcommand{\ee}{\end{equation}}
\newcommand{\bA}{\begin{align}}
\newcommand{\eA}{\end{align}}
\newcommand{\bF}{\begin{figure}}
\newcommand{\eF}{\end{figure}}

\makeatletter
\newcommand\footnoteref[1]{\protected@xdef\@thefnmark{\ref{#1}}\@footnotemark}
\makeatother

\graphicspath{{Figures/}}

\begin{document}
\title{Continuous-Variable Quantum Computing in Optical Time-Frequency Modes using Quantum Memories}
\author{Peter C. Humphreys}
\author{W. Steven Kolthammer}
\author{Joshua Nunn}
\affiliation{Clarendon Laboratory, Department of Physics, University of Oxford, OX1 3PU, United Kingdom}
\author{Marco Barbieri}
\affiliation{Clarendon Laboratory, Department of Physics, University of Oxford, OX1 3PU, United Kingdom}
\affiliation{Dipartimento di Scienze, Universitˆ degli Studi Roma Tre, Via della Vasca Navale, 84 - 00146 Rome, Italy}
\author{Animesh Datta}
\author{Ian A. Walmsley}
\affiliation{Clarendon Laboratory, Department of Physics, University of Oxford, OX1 3PU, United Kingdom}

\begin{abstract}
We develop a scheme for time-frequency encoded continuous-variable cluster-state quantum computing using quantum memories. In particular, we propose a method to produce, manipulate and measure 2D cluster states in a single spatial mode by exploiting the intrinsic time-frequency selectivity of Raman quantum memories. Time-frequency encoding enables the scheme to be extremely compact, requiring a number of memories that is a linear function of only the number of different frequencies in which the computational state is encoded, independent of its temporal duration. We therefore show that quantum memories can be a powerful component for scalable photonic quantum information processing architectures.
\end{abstract}
\maketitle

The need for ever-higher bandwidth encoding of information in optical fields has spurred innovation in modern telecommunication technologies. State of the art wavelength-division-multiplexed systems can now provide terabytes per second of capacity in a single fiber by leveraging sophisticated time-frequency (TF) encodings~\cite{Keiser1999}. In contrast, optical quantum information processing (QIP) has historically relied on spatial or polarization encodings, providing at most two quantum channels per optical spatial mode. However, as the complexity of these quantum optical experiments increases, these primitive encodings will not be sufficient. The appeal of higher density encodings for quantum optical information applications is twofold. Firstly, optical quantum communication schemes will require considerably higher bit-rates to compete with classical schemes. Secondly, for QIP protocols, the need to efficiently mediate interactions between large numbers of different modes will challenge the scalability of current encodings.

\begin{figure}[htbp]
\begin{center}
\includegraphics[width=7.5cm]{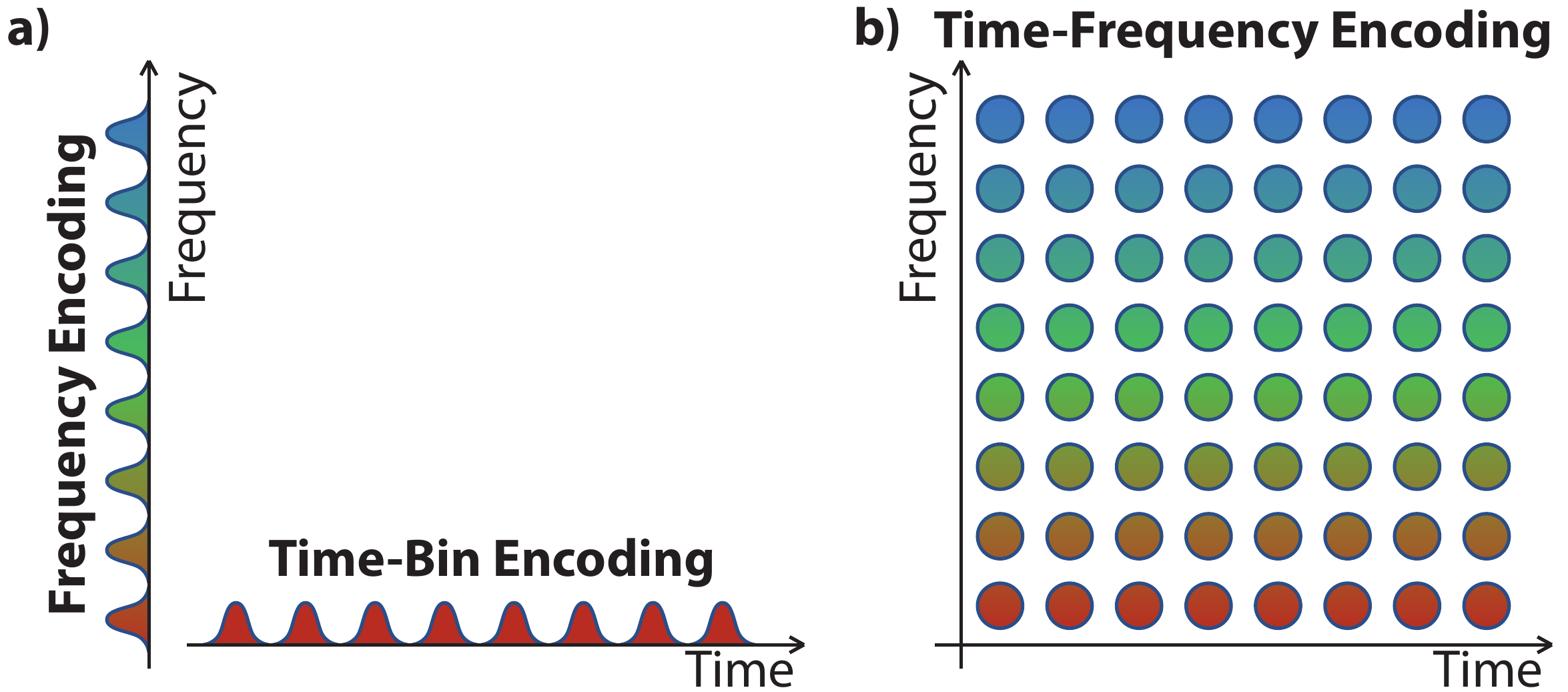}
\caption{Quantum information can be encoded in a) central-frequency modes or time bin modes. However, the maximum channel capacity can be achieved by b) tiling minimum uncertainty TF modes across the entire available encoding space.}
\label{fig:TimeFreq}
\end{center}
\end{figure}

Drawing inspiration from telecommunications, recent works have proposed methods of using time or frequency encodings to access high-dimensional Hilbert spaces~\cite{DeRiedmatten2004, Hayat2012, Nunn2013a, Nisbet-jones2013}. Further, attention has recently been drawn to the manipulation of optical quantum information in the TF domain for computation~\cite{Humphreys2013a, Campbell2013}, including the realization of large discrete and continuous-variable cluster states~\cite{Lindner2009, Menicucci2010a, Menicucci2011, Yokoyama2013, Chen2014}. These approaches can be separated into frequency encodings~\cite{Menicucci2008, Roslund2013}, in which information is encoded in different central-frequency modes with the same temporal profile, and time encodings~\cite{Soudagar2007, Menicucci2010a,  Menicucci2011, Humphreys2013a, Yokoyama2013}, in which the information is encoded in different time-bin modes with the same central frequency. Irrespective of which method is chosen, these encodings can be visualized as a one-dimensional (1D) set of modes in TF space (Fig~\ref{fig:TimeFreq}a).

These encodings can lead to a significant increase in the information available in an optical channel. However, they still use far fewer modes than the ultimate channel capacity. Consider an arbitrary mode in TF space; it is constrained by the Fourier uncertainty principle, which defines a minimum area region occupied by each mode, $\delta t \, \delta \omega \geq 1/2$. Using a two-dimensional (2D) encoding in which these minimal area plaquettes are tiled over the total available frequency bandwidth and time duration therefore achieves the maximum channel capacity~(Fig.~\ref{fig:TimeFreq}b)~\footnote{In Fig.~\ref{fig:TimeFreq}, for ease of discussion and illustration, we illustrate a set of modes with well separated time and central-frequency degrees of freedom. This modal decomposition can be made asymptotically orthogonal, however, it should be noted that it does not achieve the theoretical maximum support in time-frequency space. Modal decompositions such as Hermite-Gaussian modes are compatible with our scheme, and achieve a higher channel capacity.}. In order to take full advantage of this large potential encoding space, methods must be developed for the coherent generation, manipulation and measurement of TF information. In this Letter, we propose such a scheme for continuous-variable (CV) cluster-state computing. Although, for clarity of discussion, we specifically consider this model for quantum computing, the methods we develop should be widely applicable to many QIP tasks.

\emph{Continuous-variable quantum computing-} In CV quantum computing~\cite{Braunstein2005, Gottesman2001a, Weedbrook2012}, information is encoded in the quantum state of a field mode, often termed a quantum mode, or `qumode'. Qumodes exist in infinite-dimensional Hilbert spaces~\cite{Ferraro2005,Braunstein2005}, spanned by a continuum of orthogonal states, here denoted by $\ket{q_i}$ for $q_i \in \mathbb{R}$, where $i$ is used to label the mode. The conjugate basis is $ \ket{p_i} = \frac{1}{2\pi} \int^\infty_{-\infty} \mathrm{d}q_i \, e^{\mathrm{i} p_i q_i} \ket{q_i}$.
We define observables $\hat{q_i}$ and $\hat{p_i}$ such that $\hat{q_i}\ket{q_i} = q_i \ket{q_i}$ and $\hat{p_i}\ket{p_i} = p_i \ket{p_i}$, with  $[\hat{q}_i, \hat{q}_j] = 0$, $[\hat{q}_i, \hat{p}_j] = \mathrm{i} \, \delta_{i,j}$ taking $\hbar = 1$. Typically for optical quantum computing,  $\ket{q_i}$ represents a quadrature eigenstate of the electromagnetic field, in which case we can write $ \hat{q}_i = \frac{1}{\sqrt{2}} (\hat{a}_i + \hat{a}_i^\dag), \quad \hat{p}_i = \frac{1}{\mathrm{i} \sqrt{2}} (\hat{a}_i - \hat{a}_i^\dag)$, where $\hat{a}_i$ is the annihilation operator associated with the field mode.

Cluster-state quantum computing is an approach to the manipulation of CV quantum information using a resource `cluster state' of entangled qumodes~\cite{Menicucci2006, Gu2009,VanLoock2007}. Cluster states are often represented as a graph, with qumodes as nodes, and edges representing entanglement between qumodes (e.g. as shown in Fig.~\ref{fig:2DCluster}). Remarkably, it can be shown that Gaussian measurements of such a 2D grid of entangled qumodes can be used to implement arbitrary Gaussian operations on a set of initial qumodes~\cite{Braunstein2005}. With the additional capability to perform, deterministically, a single type of non-Gaussian operation, this approach can be shown to be sufficient to allow for universal quantum computing~\cite{Gottesman2001a}. The size of a CV quantum machine is limited by the number of qumodes across which one can create entanglement. This is a particularly significant limitation for cluster-state schemes, as these require the production of a resource state of qumodes that scales both with the number of qumode states required to encode the problem of interest and with the number of operations to be implemented. These requirements make cluster-state quantum computing an ideal candidate to benefit from the high dimensionality of time-frequency encodings.

\emph{Time non-stationary elements-} In order to be able to generate and manipulate a CV cluster state based on TF plaquettes, it is of fundamental importance to be able to implement both time stationary and non-stationary linear transformations~\cite{Brecht2014,Reddy2014,Saglamyurek2014}~\footnote{See Supplemental Material section I for further discussion of the requirement for time non-stationary transformations.}. Although such elements have been demonstrated for purely time-bin encoded~\cite{Soudagar2007, Hosseini2009} and purely frequency encoded~\cite{Huntington2004, Olislager2010, Hayat2012} quantum operations, the dual requirement for both time and frequency manipulation is more challenging to satisfy~\cite{Brecht2014,Reddy2014}. Here we propose a method for controllably producing, manipulating and measuring TF encoded quantum states by exploiting the powerful modal selectivity of Raman light-matter interactions. These interactions have been demonstrated in diverse systems including alkali atom quantum memories~\cite{Reim2012,Bao2012a,Bustard2013a,Hammerer2010, Hosseini2009}, bulk diamond~\cite{Lee2011b} and opto-mechanical resonators~\cite{Palomaki2013}.

We consider an electromagnetic field mode, with a central frequency $\omega_i$, that is temporally localized near local time $\tau_i = t - z/c$, where $z$ represents the position along the direction of propagation (restricting ourselves to one spatial mode) and $t$ is the laboratory time. We define an annihilation operator $\hat{a}_i$ for the mode. This field mode is coupled to a stationary memory mode $\hat{b}$ by an interaction Hamiltonian 
\begin{align}
\hat{H} = \, &K_{ij} \, \hat{b}^\dagger \, \hat{a}_j^\dagger \, \hat{a}_i \, + K^*_{ij} \, \hat{a}_i^\dagger \, \hat{a}_j \, \hat{b}\label{eqn:RamHam}.
\end{align}
The coupling is mediated by a control field mode with central frequency $\omega_j$, localized near $\tau_j$, with an associated annihilation operator $\hat{a}_j$. The intrinsic strength of the coupling is determined by $K_{ij}$, which is a property of the specific system considered. The creation and annihilation operators for the modes obey the standard commutation relations $[ \hat{a}_i, \hat{a}_j^\dag] = \delta_{ij}$, $[ \hat{a}_i, \hat{a}_j]  =  [ \hat{a}_i, \hat{b}] = [ \hat{a}_i, \hat{b}^\dagger] = 0$, $[\hat{b}, \hat{b}^\dag] = 1$.

This phenomenological Hamiltonian may be used to describe several useful physical systems, and is the basis of Raman quantum memories, in which an electromagnetic mode is coupled into a stationary mode, stored, and subsequently read out through appropriate application of the control field.

In the case when the control field is a bright coherent state, we can replace $\hat{a}_j$ with a c-number $\gamma_j$ representing the field amplitude. In this case the effective Hamiltonian between the electromagnetic and stationary modes takes the form of a beam splitter (BS) interaction~\footnote{See Supplemental Material section II, which includes Ref.~\cite{Ferraro2005}, for a detailed discussion on the use of the BS and TMS Hamiltonians for modal manipulations, including a discussion of how to incorporate arbitrary phase shifts into a sequence of operations.}
\begin{align}
\hat{H}_{\mathrm{BS}} = \, & \gamma_j  K_{ij} \, \hat{b}^\dagger \, \hat{a}_i \, + \gamma_j^* K^*_{ij} \, \hat{a}_i^\dagger \, \hat{b}\label{eqn:BS}.
\end{align}

Crucially, since the coupling, via $\gamma_j(\omega_j, \tau_j)$, is a function of $\omega_j$ and $\tau_j$, the control field can be adjusted in order to coherently couple different TF modes with the stationary memory mode~\cite{Hammerer2010,Campbell2013,Saglamyurek2014}. This TF dependent beam splitter is exactly the time non-stationary element that is required in order to carry out TF linear optical quantum computing. This was recognized in a recent paper on the implementation of linear unitaries between different central-frequency modes using quantum memories~\cite{Campbell2013}. In our scheme, we use this interaction to couple arbitrary TF plaquettes (Fig.~\ref{fig:TimeFreq}b). With this element, it is therefore possible to implement TF versions of any linear optical quantum computing scheme.
\begin{figure}
\begin{center}
\includegraphics[width=8cm]{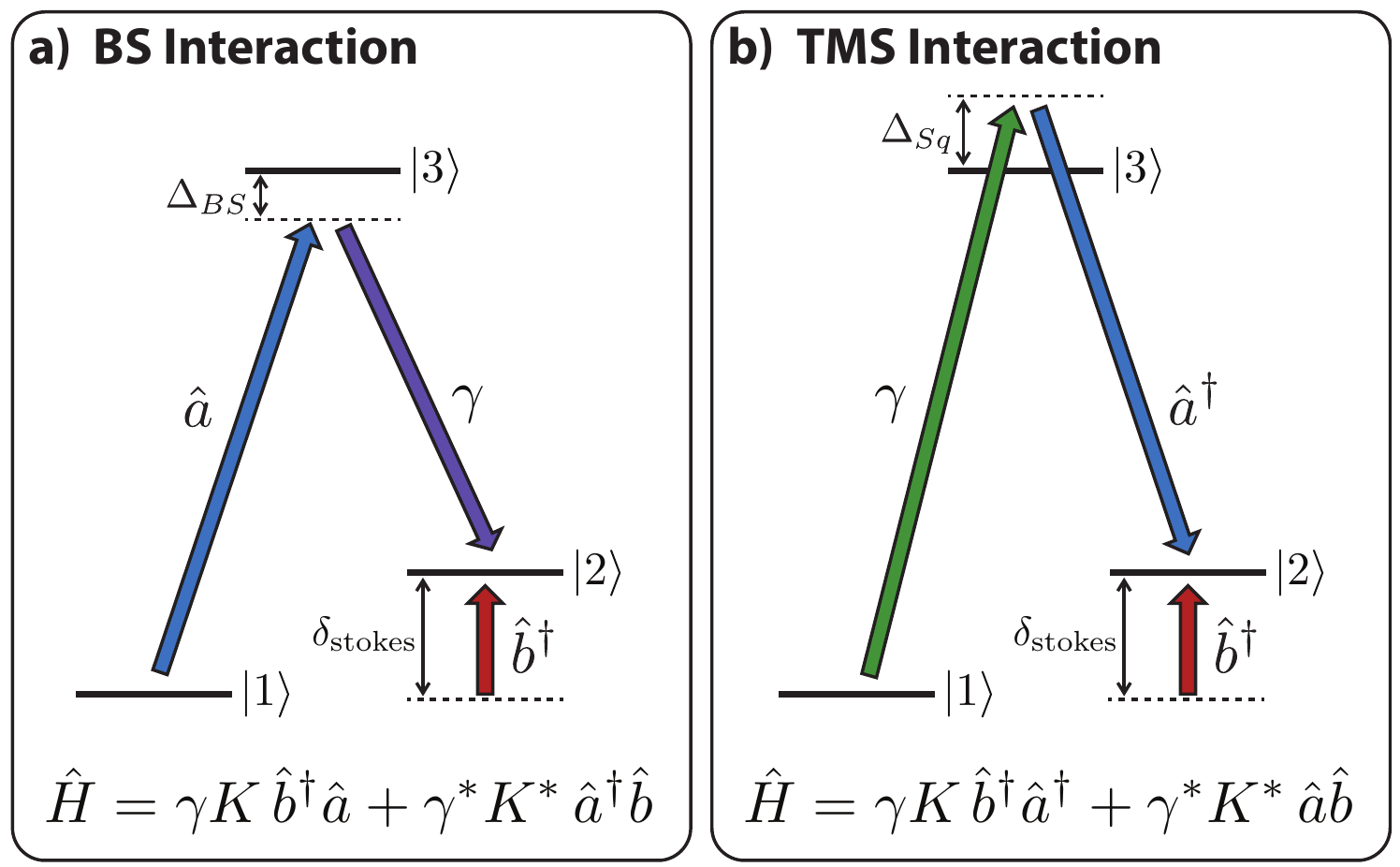}
\caption{An example off-resonant Raman memory scheme based on a three-level system. An electromagnetic mode $\hat{a}$ is coupled to the transition $\hat{b}$ between levels $\ket{1}$ and $\ket{2}$ through a Raman transition to a virtual energy level detuned by $\Delta$ from $\ket{3}$. By adjusting the frequency of the control field-mode $\gamma$, both a) beam splitter interactions and b) two-mode squeezing interactions can be generated between the modes. The BS operation can realise read in/out from the memory.}
\label{fig:Raman}
\end{center}
\end{figure}

However, for continuous-variable quantum computing, the utility of Raman quantum memories extends beyond linear mode manipulations. If, in Eq.~\ref{eqn:RamHam}, the classical limit of the $\hat{a}_i$ field-mode is instead taken, replacing $\hat{a}_i$ with $\gamma_i$, two-mode squeezing (TMS) can be generated~\footnotemark[3] between the $\hat{a}_j$ mode and the stationary $\hat{b}$ mode~\cite{Datta2012, Chakhmakhchyan2013, Hammerer2010}
\begin{align}
\hat{H}_{\mathrm{TMS}} = \, & \gamma_i K_{ij} \, \hat{b}^\dagger \, \hat{a}^\dagger_j  \, + \gamma_i^* K_{ij}^* \, \hat{a}_j \, \hat{b}.
\end{align}
In this case, the electromagnetic mode that is coupled to the memory has changed. However, in many physical systems, it is possible to adjust the frequency of the classical field used so that the electromagnetic mode coupled is the same for BS and TMS operations. This provides us with a universal time non-stationary element, that can be used to implement BS and TMS operations between a stationary mode and near-arbitrary TF modes (Fig.~\ref{fig:Raman}). As we show below, this enables the full generation and manipulation of TF CV cluster states.

\emph{CV cluster state generation-} In the canonical model of CV cluster-state computing, the resource state is composed of a number of qumodes, each prepared in the $\{p_i = 0\}$ eigenstate $\ket{0_{p_i}} = \int_{-\infty}^{\infty}\ket{q_i} dq_i$. This is the infinitely-squeezed limit of a single-mode squeezed vacuum state, and therefore can only be prepared approximately~\cite{Menicucci2014}. These qumodes must subsequently be entangled with each other as required to create a given cluster state geometry.

\begin{figure}[h]
\begin{center}
\includegraphics[width=5cm]{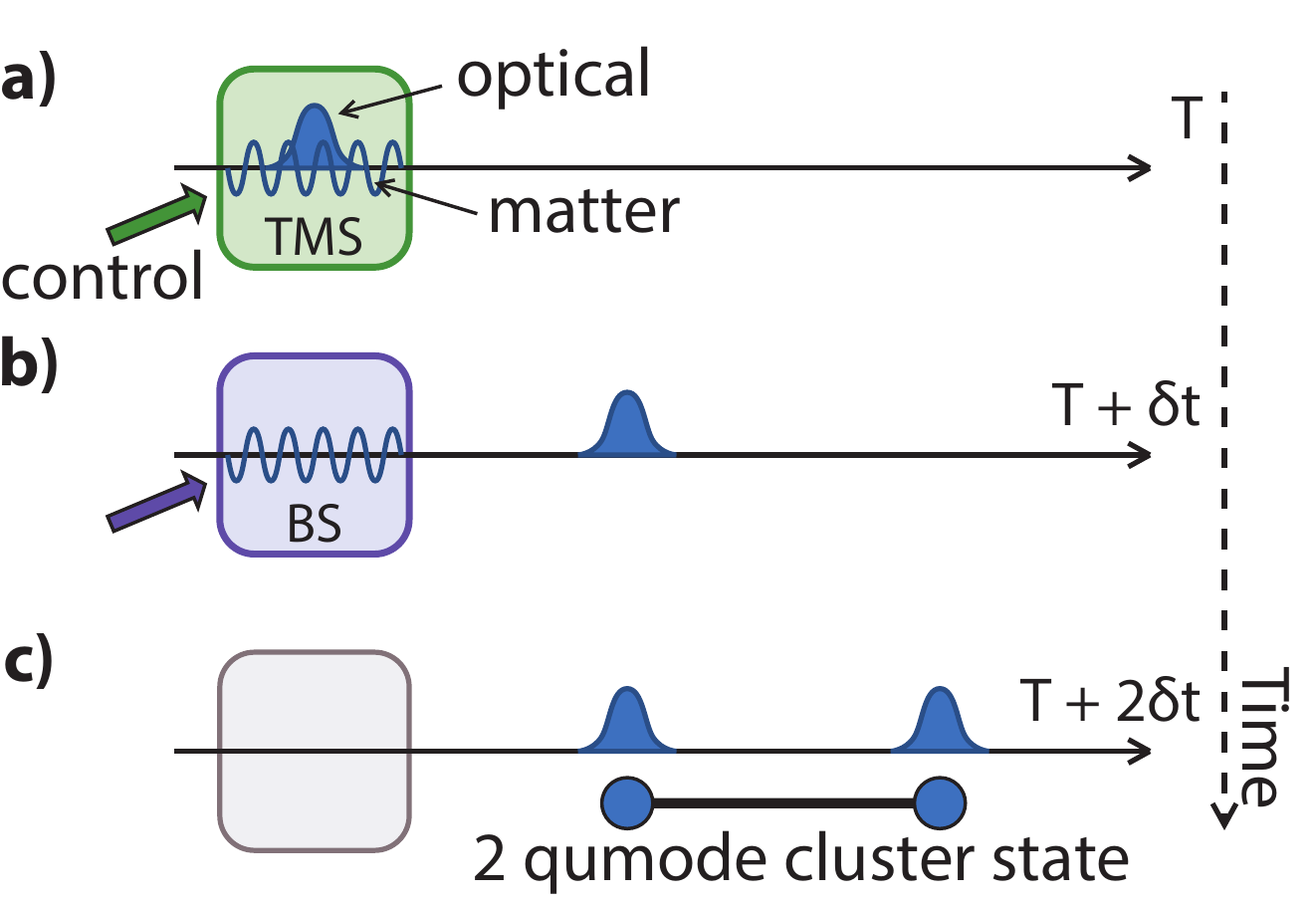}
\caption{a) Initial time-bin two-qumode cluster states can be produced by using a Raman memory to create a TMS interaction between electromagnetic and memory vacuum modes. b) A BS interaction is then used to read out the memory qumode into another TF plaquette, resulting in c) a flying 2-TF mode cluster state}
\label{fig:basicCluster}
\end{center}
\end{figure}

\emph{Two-qumode cluster-} As a first step towards a full cluster state, we demonstrate how to use the TMS interaction to create two-qumode TF encoded cluster states. These will function as primitives for our full state. To do this, we note \cite{VanLoock2007} that a two-mode canonical cluster state provides the same correlations as a two-mode squeezed state under a rotation of the quadratures of one of the modes, so that $q_i \rightarrow p_i$ and $p_i \rightarrow -q_i$. Therefore application of the TMS interaction between a vacuum electromagnetic mode and a memory initialized in its ground state will produce a qumode in the memory entangled with a qumode in the field (Fig.~\ref{fig:basicCluster}). A subsequent BS interaction can be used to `read out' the memory qumode by coupling it into another TF plaquette. This produces a two-qumode TF cluster state~\footnote{See Supplemental Material section IV, which includes Refs.~\cite{Hosseini2011a,Datta2012, Menicucci2014,Marian2012,Barnum1996,Gottesman2001a,Ferrini2014,Kieling2007}, for a discussion on the impact of imperfect memory operations on cluster state generation.}
 (equivalent to an EPR state~\cite{Braunstein2005}).

\begin{figure}[htbp]
\begin{center}
\includegraphics[width=6.5cm]{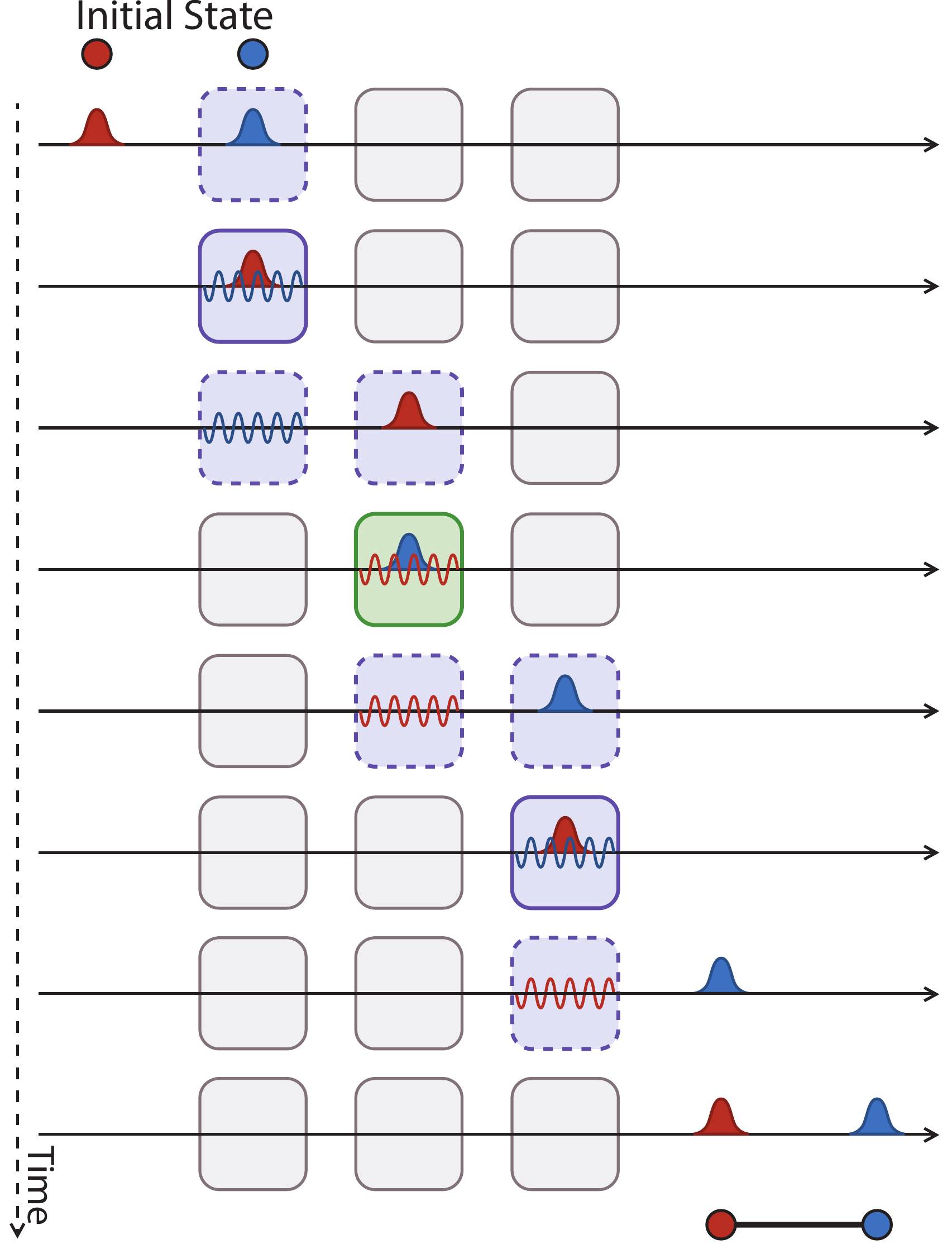}
\caption{Three separate two-qumode operations are required to create a CZ gate, in the sequence BS, TMS, BS. Here we show the full protocol needed to implement such a gate between two qumodes encoded in different time-frequency plaquettes (control fields are omitted for clarity). This can be achieved by using a set of three quantum memories. Only three operations are used to implement the CZ gate (solid outlines), the other BS operations are simply required in order to maintain the temporal separation of the two qumodes (dashed outlines). The two qumodes are shaded in different colors for clarity, but do not necessarily have different central-frequencies.}
\label{fig:CZOperation}
\end{center}
\end{figure}

\begin{figure}[htbp]
\begin{center}
\includegraphics[width=8.5cm]{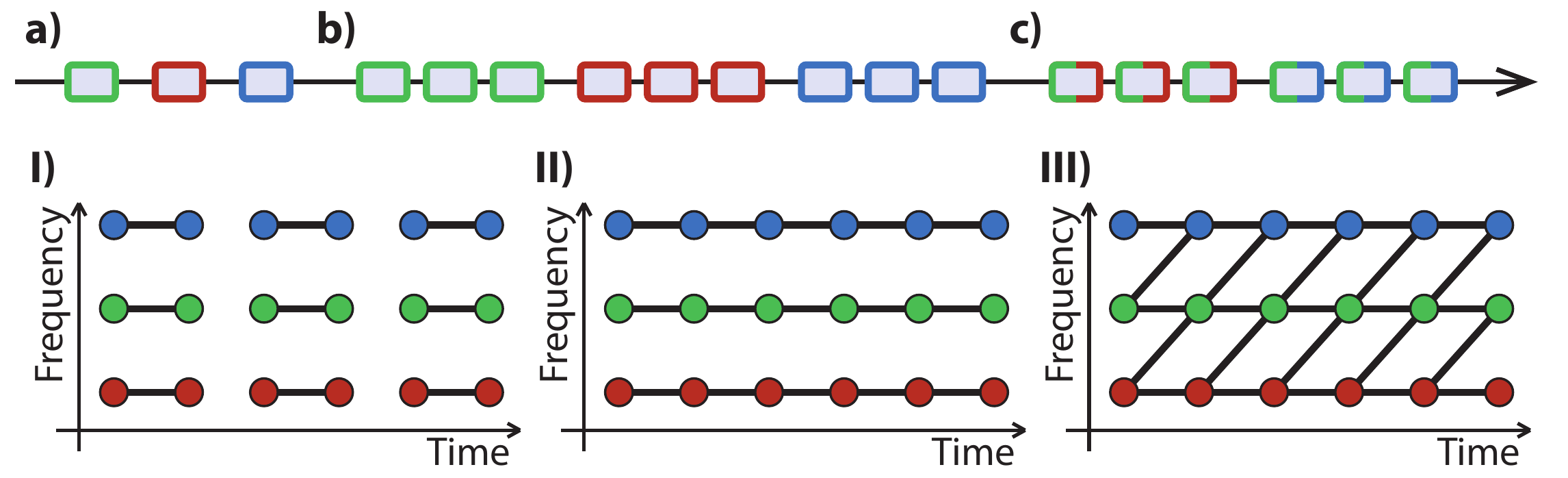}
\caption{A linear chain of $7d -3$ quantum memories can generate 2D cluster states encoded in $d$ different central-frequency modes and a large number of time-bin modes. a) $d$ quantum memories are used to create two-qumode clusters encoded in plaquettes tiled across the available TF space, as shown in I) (colors are used to denote different central-frequency plaquettes, with the color of the quantum memory corresponding to the central-frequency that it operates upon). b) Entanglement links are generated between two qumode clusters with the same central frequency using CZ interactions, as shown in II). This requires a further $3d$ memories. c) Entanglement is generated between qumodes with different central-frequencies, requiring a final set of $3(d-1)$ memories. This produces the final 2D cluster state shown in III).}
\label{fig:2DCluster}
\end{center}
\end{figure}

\emph{2D cluster-} After creation, these two-qumode clusters must then be linked to create a 2D cluster state. The canonical method for generating these links is to apply a so-called controlled-Z (CZ) gate, which implements the unitary operation $e^{\mathrm{i} \hat{q}_i \hat{q}_j}$~\cite{Menicucci2006, Ukai2011a}. For this purpose, the TMS operation can not substitute the CZ gate, as the qumodes are not initially in the vacuum. The ability to implement the CZ gate also enables other cluster state computing schemes, such as the memory based cluster-state scheme described in~\cite{Roncaglia2011}. A CZ gate can be realized in our scheme through a sequence of a BS interaction, a TMS interaction, and an additional BS interaction between two qumodes~\footnote{See Supplemental Material section III, which includes Refs.~\cite{Braunstein2005a, Ferraro2005}, for details on the implementation of a CZ gate in our scheme.}. A straightforward method for implementing this uses three successive memories~\footnotemark[4], as shown in Fig.~\ref{fig:CZOperation}. It should be noted that it is also possible to use only BS operations to link qumodes~\cite{Menicucci2011, Yokoyama2013} although, in this case, corrective displacement operations must be physically implemented on each qumode between measurements. An alternative approach to creating cluster states uses only vacuum squeezing followed by linear interactions~\cite{VanLoock2007}. However, the number of linear interactions required scales quadratically with the number of qumodes in the cluster.

In order to create a full 2D cluster state, we exploit the precise modal selectivity of the Raman memory to controllably address different TF plaquettes. This allows us to implement the CZ operations required to entangle neighboring initial two-qumode cluster states (Fig.~\ref{fig:2DCluster}), creating the desired overall cluster topology. The modes within a time-bin that are not being addressed by a given memory will pass straight through, allowing each link to be created completely independently.

In this way, a linear chain of only $7d -3$ quantum memories is sufficient to generate 2D cluster states encoded in $d$ different central-frequency modes. The cluster state can be of significant temporal duration, since a given qumode is only stored in a memory for the length of time $\delta t$ required to implement a BS or TMS operation. This requires that the coherence time of the memory $t_{\mathrm{mem}} \gg \delta t$. However, $t_{\mathrm{mem}}$  can be significantly less than the temporal extent of the entire cluster, so that $t_{\mathrm{mem}} \ll n \, \delta t$, where $n$ is the number of time-bin modes. For a given $\delta t$ and therefore $\delta \omega$, the number of different central-frequency modes $d$ is then constrained by the range of frequencies $\delta_{\text{full}}$ that the memory can access. For example, in memories based on alkali atoms we require $\delta_{\text{full}} \ll \delta_{\text{stokes}}$ (Fig.~\ref{fig:Raman}) in order to suppress four-wave-mixing noise~\cite{Reim2011}. Therefore the maximum $d$ is on the order of $\delta_{\text{full}} / \delta \omega \approx \delta_{\text{full}} \, \delta t$, which is the standard time-bandwidth-product figure of merit for quantum memories. Time-bandwidth-products upwards of 1000 have been demonstrated~\cite{Reim2011,Bao2012a}.

\begin{figure}[h]
\begin{center}
\includegraphics[width=6cm]{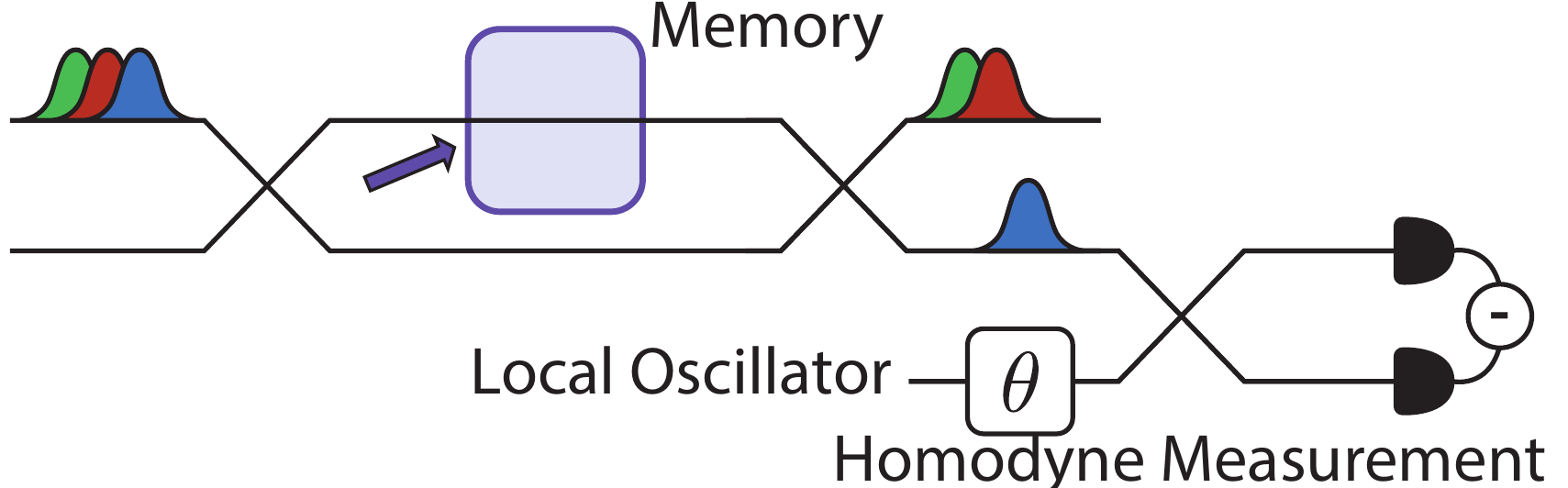}
\caption{Measurement of a frequency qumode by using a quantum memory embedded in a balanced Mach-Zehnder interferometer. A BS operation is used to apply a $\pi$ phase to the qumode to be measured, changing the output port it exits from. An arbitrary quadrature of the qumode can then be measured by using a local oscillator to carry out homodyne detection. This scheme may be implemented in a signal spatial mode by using appropriate polarization elements.}
\label{fig:Meas}
\end{center}
\end{figure}

\emph{Measurements and Non-Gaussian Operations-} TF cluster states can be used to perform quantum information processing tasks. This requires the measurement of an appropriate field quadrature for each qumode in the cluster. Since different central-frequency qumodes are co-propagating in the same time-bin, a method of measuring each of these modes separately is needed. For this, we can again exploit  the TF selectivity of a quantum memory. Consider embedding a memory in one arm of a balanced Mach-Zehnder interferometer, with path lengths adjusted so that the qumodes are mapped back into their original mode (Fig.~\ref{fig:Meas}a). A BS operation can be applied to one frequency qumode that is sufficiently strong to read it in and then instantaneously read it out of the memory. It can be easily shown that this will create a $\pi$ phase shift in this mode and therefore swap the output port from which it will exit. Homodyne detection can then be carried out on this qumode without disturbing the others.


In order to carry out universal quantum computation, non-linear operations or non-linear resource states are also needed~\cite{Lloyd1999,Raussendorf2003}. Direct implementation of non-linear operations is experimentally challenging~\cite{Peyronel2012}, therefore most proposed schemes have been based on measurement-induced non-linearities~\cite{Gottesman2001a, Wenger2004}. Our method naturally provides a convenient platform for probabilistic non-Gaussian state generation schemes, since the qumodes are already stored in quantum memories at points of the protocol. The interval between storage and read-out therefore provides a period in which multiple attempts can be made at generating suitable non-linear states. The use of quantum memories to increase the success rate of photon subtraction has been studied in recent papers on entanglement distillation~\cite{Datta2012, Chakhmakhchyan2013}.

\emph{Conclusions-}We have proposed novel active elements for TF linear optical quantum computing based on Raman quantum memories. These elements are able to implement time non-stationary beam splitter and two mode squeezing interactions, providing a versatile toolbox for state generation, manipulation and measurement. Our results suggest a useful route towards maximize the space of encodings available for electromagnetic modes, and could also be combined with other degrees of freedom, including polarization and spatial encodings, in order to obtain further increases in dimensionality. This may enable innovative approaches to QIP~\cite{Raussendorf2005a,Barreiro2005}.

\emph{Acknowledgments- }We thank G. Ferrini, N. Treps, D. Saunders, B. Metcalf and J. Spring for helpful discussions. This work was supported by the EPSRC(EP/K034480/1, EP/H03031X/1, EP/H000178/1), the EC project SIQS, the Royal Society, and the AFOSR EOARD. WSK is supported by EC Marie Curie fellowship (PIEF-GA-2012-331859). AD is supported by the EPSRC (EP/K04057X/1). MB is funded by a Rita Levi-Montalcini contract of MIUR.

\bibliography{CVPapers}

\clearpage
\begin{center} \large{Supplemental Material} \end{center}

\section{Requirement for time non-stationary coupling elements}
A time-frequency (TF) manipulation scheme requires an operation that can displace a qumode in TF space from $(\omega_i, t_i)$ to $(\omega_j, t_j)$. This can be implemented by first translating in frequency, and then in time. In the following we briefly outline an argument showing that time non-stationary interactions are necessary in order to couple different frequency modes.

For a linear interaction with an optical mode, the output field $E_{out}$ is related to the input mode $E_{in}$ by
\begin{align}
E_{out}(t) = \int^{\infty}_{-\infty} R(t,t')E_{in}(t') \mathrm{d}t'.
\end{align}

If the response function $R(t,t')$ is time-stationary, it can only depend on the time difference $t' - t$, giving
\begin{align}
E_{out}(t) = \int^{\infty}_{-\infty} R(t' - t)E_{in}(t') \mathrm{d}t'.
\end{align}

This integral has the form of a convolution, and therefore the Fourier transform gives
\begin{align}
E_{out}(\omega) = \chi(\omega)E_{in}(\omega),
\end{align}
where the linear susceptibility $\chi(\omega)$ is the Fourier transform of $R(t)$. This only couples each frequency to itself.

\section{Raman memory operations}

As discussed in the main text, it is possible to implement two different effective Hamiltonians between a stationary memory mode and an electromagnetic field mode.

\subsection{Beamsplitter operation}

The beamsplitter (BS) operation has the Hamiltonian
\begin{align}
\hat{H}_{\mathrm{BS}} = \, & \gamma_j  K_{ij} \, \hat{b}^\dagger \, \hat{a}_i \, + \gamma_j^* K^*_{ij} \, \hat{a}_i^\dagger \, \hat{b}\label{eqn:BS}.
\end{align}
where $\hat{b}$ represents the stationary mode, and $\hat{a}_i$ a given field mode.

In the Heisenberg picture, it can be shown~\cite{Ferraro2005} that this Hamiltonian induces a mode transformation
\begin{align}
\begin{pmatrix}
\hat{a}_i' \\
\hat{b}'
\end{pmatrix} =
\begin{pmatrix}
\cos{\phi t} & e^{-\mathrm{i} \theta} \sin{\phi t} \\
- e^{\mathrm{i} \theta} \sin{\phi t} & \cos{\phi t}
\end{pmatrix}
\begin{pmatrix}
\hat{a}_i \\
\hat{b}
\end{pmatrix}
\end{align}
where we have made the substitution $\phi \, e^{\mathrm{i} \theta} = \mathrm{i} \, \gamma_i K_{ij}/ \hbar$, and $\hat{a}_i'$, $\hat{b}'$ represent the modes after the application of the Hamiltonian for a time $t$. Without loss of generality, we can incorporate this $t$ dependence into $\phi$.

It is useful to note that we can rewrite this matrix equation in the following form
\begin{align}
\begin{pmatrix}
\hat{a}_i' \\
\hat{b}'
\end{pmatrix} =
\begin{pmatrix}
1 & 0 \\
0 & e^{\mathrm{i} \, \theta}
\end{pmatrix}
\begin{pmatrix}
\cos{\phi} & \sin{\phi} \\
-\sin{\phi} & \cos{\phi}
\end{pmatrix}
\begin{pmatrix}
1 & 0 \\
0 & e^{-\mathrm{i} \, \theta}
\end{pmatrix}
\begin{pmatrix}
\hat{a}_i \\
\hat{b}
\end{pmatrix}
\end{align}

Written in this way, it is evident that this Hamiltonian produces a transformation equivalent to the application of a phase shift of $-\theta$ to mode $\hat{b}$, followed by a `standard' (real coefficients) BS operation, and a final phase shift of $\theta$ to the same mode (Fig~\ref{fig:BSandTMSdecomp}a).

\subsection{Two mode squeezing operations}

It is possible to carry out a similar procedure for the two-mode squeezing (TMS) operation, with associated Hamiltonian
\begin{align}
\hat{H}_{\mathrm{TMS}} = \, & \gamma_i K_{ij} \, \hat{b}^\dagger \, \hat{a}^\dagger_j  \, + \gamma_i^* K_{ij}^* \, \hat{a}_j \, \hat{b}.
\end{align}

In the Heisenberg picture, with the substitution $r \, e^{\mathrm{i} \psi} = -\mathrm{i} \, \gamma_i K_{ij} t/ \hbar$, this operation induces a mode transformation
\begin{align}
\begin{pmatrix}
\hat{a}_j' \\
\hat{b}'
\end{pmatrix} =
\begin{pmatrix}
\mu & 0 \\
0 & \mu
\end{pmatrix}
\begin{pmatrix}
\hat{a}_j \\
\hat{b}
\end{pmatrix}
+
\begin{pmatrix}
0 & e^{\mathrm{i} \psi} \nu \\
e^{\mathrm{i} \psi} \nu & 0
\end{pmatrix}
\begin{pmatrix}
\hat{a}^\dagger_j \\
\hat{b}^\dagger
\end{pmatrix}
\end{align}
where $\mu = \cosh{r}$ and $\nu = \sinh{r}$.

This can be rewritten in the form

\begin{align}
\begin{pmatrix}
\hat{a}_j' \\
\hat{b}'
\end{pmatrix} =&
\begin{pmatrix}
1 & 0 \\
0 & e^{\mathrm{i} \psi}
\end{pmatrix}
\begin{pmatrix}
\mu & 0 \\
0 & \mu
\end{pmatrix}
\begin{pmatrix}
1 & 0 \\
0 & e^{-\mathrm{i} \psi}
\end{pmatrix}
\begin{pmatrix}
\hat{a}_j \\
\hat{b}
\end{pmatrix}\notag\\
&+
\begin{pmatrix}
1 & 0 \\
0 & e^{\mathrm{i} \psi}
\end{pmatrix}
\begin{pmatrix}
0 & \nu \\
\nu & 0
\end{pmatrix}
\begin{pmatrix}
1 & 0 \\
0 & e^{\mathrm{i} \psi}
\end{pmatrix}
\begin{pmatrix}
\hat{a}^\dagger_j \\
\hat{b}^\dagger
\end{pmatrix}
\end{align}.

\begin{figure}[htbp]
\begin{center}
\includegraphics[width=4cm]{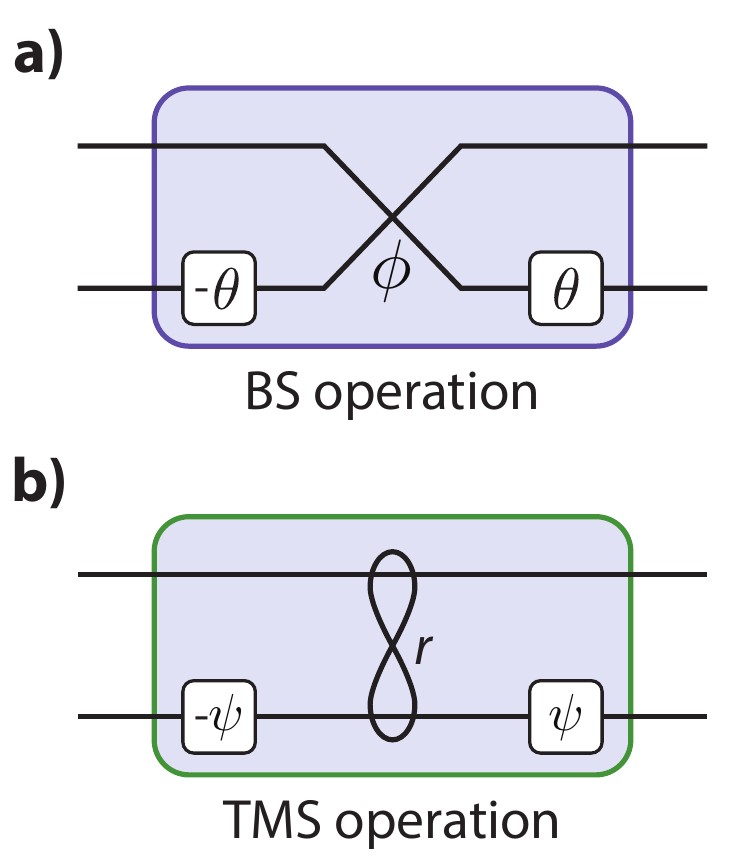}
\caption{a) The Raman memory BS operation can be decomposed into a phase shift of $-\theta$ to mode $\hat{b}$, followed by a standard BS operation parameterised by $\phi$, and a final phase shift of $\theta$ to the same mode. The coefficients $\theta$ and $\phi$ are determined by the control field applied. b) The TMS operation can be similarly decomposed into a phase shift of $-\psi$ to mode $\hat{b}$, followed by a standard TMS operation parameterised by $r$, and a final phase shift of $\psi$ to the same mode.}
\label{fig:BSandTMSdecomp}
\end{center}
\end{figure}

This Hamiltonian therefore produces a transformation equivalent to the application of a phase shift of $-\psi$ to mode $\hat{b}$, followed by a `standard' (real coefficients) TMS operation, and then a final phase shift of $\psi$ to the same mode (Fig~\ref{fig:BSandTMSdecomp}b).

\subsection{Arbitrary operations}

\begin{figure}[htbp]
\begin{center}
\includegraphics[width=8.5cm]{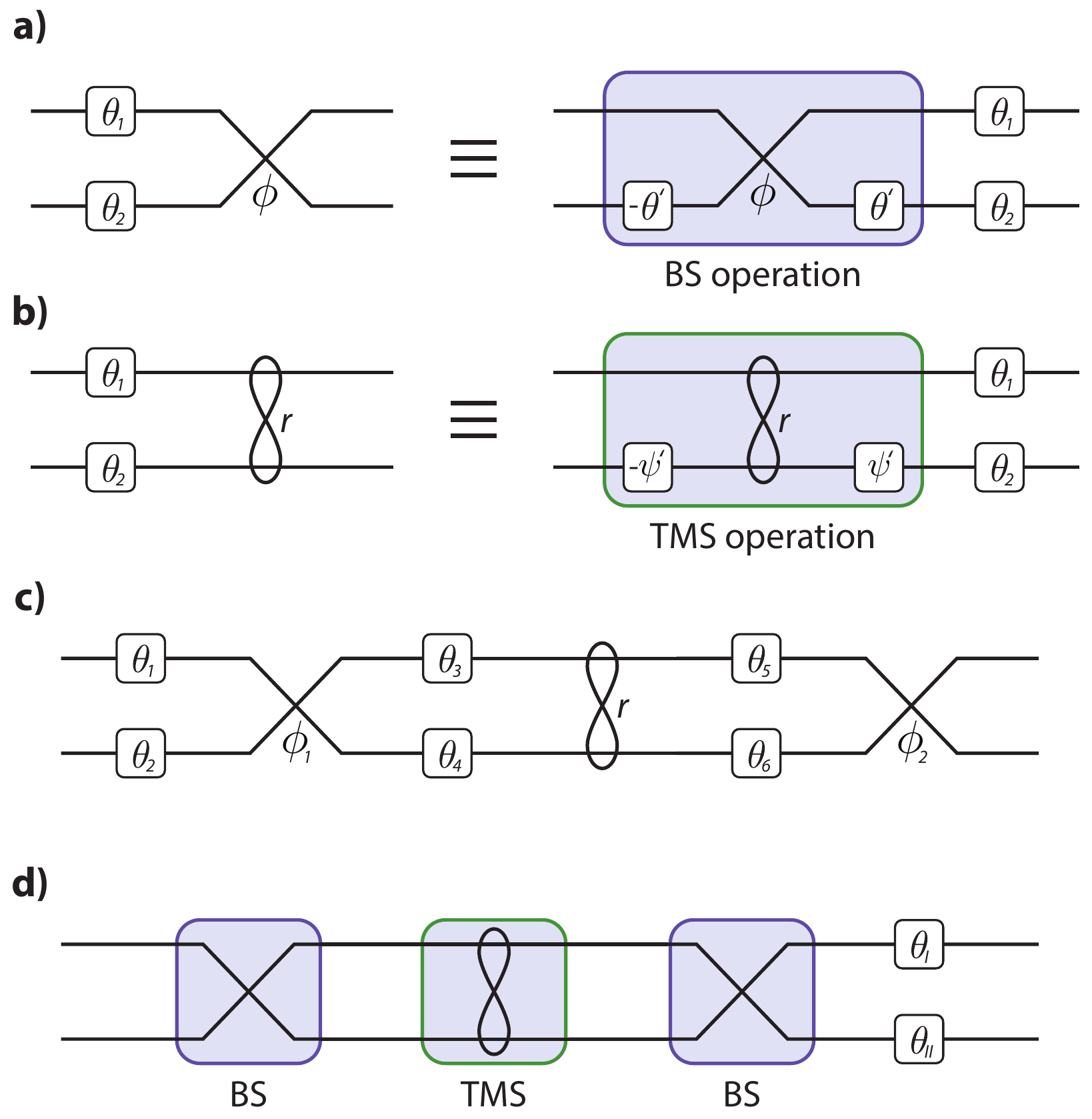}
\caption{a) A general two mode interaction consisting of the application of a phase to each mode, followed by a standard BS transformation can be shown to be equivalent to a suitable Raman BS operation (with $\theta' = (\theta_1 - \theta_2)$) followed by the same phase shifts. b) Similarly, a two mode interaction consisting of the application of a phase to each mode, followed by a standard TMS transformation can be shown to be equivalent to a suitable Raman TMS operation (with $\psi' = - (\theta_1 + \theta_2)$) followed by the same phase shifts. c) Using these equivalences, an arbitrary string of phase shifts, BS and TMS operations can be implemented using Raman memories. A final corrective phase shift for each mode can simply be incorporated into a change of measurement basis when the mode is read out.}
\label{fig:BSandTMSgenMan}
\end{center}
\end{figure}

More generally, for arbitrary linear optical manipulations, it is desirable to be able to implement transformations consisting of the application of an arbitrary phase to each mode, followed by either a standard BS or TMS squeezing operation, as shown in Figs~\ref{fig:BSandTMSgenMan}a~\&~\ref{fig:BSandTMSgenMan}b.

For the BS operation, this transformation corresponds to
\begin{align}
\begin{pmatrix}
\hat{a}_i' \\
\hat{b}'
\end{pmatrix} =
\begin{pmatrix}
\cos{\phi} & \sin{\phi} \\
-\sin{\phi} & \cos{\phi}
\end{pmatrix}
\begin{pmatrix}
e^{\mathrm{i} \, \theta_1} & 0 \\
0 & e^{\mathrm{i} \, \theta_2}
\end{pmatrix}
\begin{pmatrix}
\hat{a}_i \\
\hat{b}
\end{pmatrix}
\end{align}

It can be seen that an equivalent transformation is given by
\begin{align}
\begin{pmatrix}
\hat{a}_i' \\
\hat{b}'
\end{pmatrix} =&
\begin{pmatrix}
e^{\mathrm{i} \, \theta_1} & 0 \\
0 & e^{\mathrm{i} \, \theta_2}
\end{pmatrix}\notag\\
&*\begin{pmatrix}
\cos{\phi} & e^{-\mathrm{i} \, (\theta_1 - \theta_2)} \sin{\phi} \\
-e^{\mathrm{i} \, (\theta_1 - \theta_2)} \sin{\phi} & \cos{\phi}
\end{pmatrix}
\begin{pmatrix}
\hat{a}_i \\
\hat{b}
\end{pmatrix}
\end{align}

Therefore, by adjusting the phase of the control field so that $\theta' = (\theta_1 - \theta_2)$, it is possible to move the phase shifts to after the BS operation.

Similarly, for the TMS squeezing operation, the transformation in Fig~\ref{fig:BSandTMSgenMan}b corresponds to
\begin{align}
\begin{pmatrix}
\hat{a}_j' \\
\hat{b}'
\end{pmatrix} =&
\begin{pmatrix}
\mu & 0 \\
0 & \mu
\end{pmatrix}
\begin{pmatrix}
e^{\mathrm{i} \, \theta_1} & 0 \\
0 & e^{\mathrm{i} \, \theta_2}
\end{pmatrix}
\begin{pmatrix}
\hat{a}_j \\
\hat{b}
\end{pmatrix}\notag\\
&+
\begin{pmatrix}
0 & \nu \\
\nu & 0
\end{pmatrix}
\begin{pmatrix}
e^{-\mathrm{i} \, \theta_1} & 0 \\
0 & e^{-\mathrm{i} \, \theta_2}
\end{pmatrix}
\begin{pmatrix}
\hat{a}^\dagger_j \\
\hat{b}^\dagger
\end{pmatrix}
\end{align}

An equivalent transformation with the phase shifts moved to after the TMS operation is given by
\begin{align}
\begin{pmatrix}
\hat{a}_j' \\
\hat{b}'
\end{pmatrix} =&
\begin{pmatrix}
e^{\mathrm{i} \, \theta_1} & 0 \\
0 & e^{\mathrm{i} \, \theta_2}
\end{pmatrix}
\begin{pmatrix}
\mu & 0 \\
0 & \mu
\end{pmatrix}
\begin{pmatrix}
\hat{a}_j \\
\hat{b}
\end{pmatrix}\notag\\
&+
\begin{pmatrix}
e^{\mathrm{i} \, \theta_1} & 0 \\
0 & e^{\mathrm{i} \, \theta_2}
\end{pmatrix}
\begin{pmatrix}
0 & e^{- \mathrm{i} \, (\theta_2 + \theta_1)} \nu \\
e^{- \mathrm{i} \, (\theta_2 + \theta_1)} \nu & 0
\end{pmatrix}
\begin{pmatrix}
\hat{a}^\dagger_j \\
\hat{b}^\dagger
\end{pmatrix}
\end{align}

This can be achieved by adjusting the phase of the control field so that $\psi' = - (\theta_1 + \theta_2)$.

In this way an arbitrary string of BS operations, TMS operations and phase shifts (Fig~\ref{fig:BSandTMSgenMan}c) can be implemented using the Raman memory Hamiltonians. As shown in Fig~\ref{fig:BSandTMSgenMan}d, this is achieved by moving the desired phases through each operation to the end of the chain. The final necessary corrective phases $\theta_I$ and $\theta_{II}$ can then simply be incorporated into a change of measurement basis when each mode is finally read out.

\FloatBarrier

\section{Implementing a controlled-Z interaction using Raman memories}
For our cluster state generation scheme, it is useful to be able to directly generate a controlled-Z (CZ) operation $e^{\mathrm{i} \hat{q}_i \hat{q}_j}$ between two qumodes. This allows us to link different clusters in order to create a 2D cluster of the required size. Here we show that this can be achieved using three Raman memory interactions between an electromagnetic qumode and a memory qumode.

Following the treatment in~\cite{Braunstein2005a}, we note that all Gaussian operations can be represented as a general linear unitary Bogoliubov transformation of the form
\be
\hat{b}_j = \sum_k (A_{jk} \hat{a}_k + B_{jk} \hat{a}^\dag_k)
\ee
where $\hat{a}_k$, $\hat{b}_j$ represent annihilation operators for input mode $k$ and output mode $j$ respectively.

It can be shown~\cite{Braunstein2005a} that the matrices $A$ and $B$ can be decomposed via the Bloch-Messiah reduction into
\be
A = U A_D V^\dag, \qquad B = U B_D V^T
\ee
where $U$ and $V$ are unitary matrices and $A_D$ and $B_D$ are a pair of non-negative diagonal matrices such that $A_D^2 - B_D^2 = \mathds{1}$, with $\mathds{1}$ the identity matrix. The corollary of this is that any Gaussian operation can be decomposed into a linear interferometer, followed by parallel single-mode squeezing operations, and finally another linear interferometer.

Using this decomposition, the Bloch-Messiah reduction for the CZ Hamiltonian is given by~\footnote{The discrepancy with~\cite{Braunstein2005a} is due to a slight difference in the definition of the CZ Hamiltonian}
\begin{align}
U &= \begin{pmatrix}
\sin \phi & - \mathrm{i} \cos \phi \\
\mathrm{i} \cos \phi & - \sin \phi
\end{pmatrix},\quad
V = \begin{pmatrix}
\cos \phi & - \mathrm{i} \sin \phi \\
\mathrm{i} \sin \phi & - \cos \phi
\end{pmatrix},\notag\\
A_D &= \begin{pmatrix}
\frac{\sqrt{5}}{2} & 0 \\
0 & \frac{\sqrt{5}}{2}
\end{pmatrix},\quad
B_D = \begin{pmatrix}
\frac{1}{2} & 0 \\
0 & \frac{1}{2}
\end{pmatrix}\label{eqn:BlochMessiah}
\end{align}
where $\phi=\frac{1}{2} \sin^{-1}(2/\sqrt{5}) \approx 31.72^\circ$.

This cannot be directly implemented using Raman quantum memories, as they can only implement TMS and BS operations. However, since the squeezing is the same for each mode, it is possible to substitute a TMS operation and suitable linear optical transformations for the single mode squeezing~\cite{Ferraro2005}. A standard TMS operation, parametrised by $\mu$ and $\nu$, is given by
\begin{align}
A_{TMS} = \begin{pmatrix}
\mu & 0 \\
0 & \mu
\end{pmatrix}, \quad
B_{TMS} = \begin{pmatrix}
0 & \nu \\
\nu & 0 \\
\end{pmatrix}.
\end{align}

Applying a 50:50 beam splitter operation and $\pi/2$ phase transformation before and after the squeezing operation, we find that
\begin{align}
A' &=
\begin{pmatrix}
\frac{1}{\sqrt{2}} & \frac{1}{\sqrt{2}} \\
\frac{ \mathrm{i}}{\sqrt{2}} & -\frac{ \mathrm{i}}{\sqrt{2}}
\end{pmatrix}
\begin{pmatrix}
\mu & 0 \\
0 & \mu
\end{pmatrix}
\begin{pmatrix}
\frac{1}{\sqrt{2}} & -\frac{ \mathrm{i}}{\sqrt{2}} \\
\frac{1}{\sqrt{2}} & \frac{ \mathrm{i}}{\sqrt{2}}
\end{pmatrix}
\notag\\
&= \begin{pmatrix}
\mu & 0 \\
0 & \mu
\end{pmatrix} = A_D \notag\\
B' &=
\begin{pmatrix}
\frac{1}{\sqrt{2}} & \frac{1}{\sqrt{2}} \\
\frac{ \mathrm{i}}{\sqrt{2}} & -\frac{ \mathrm{i}}{\sqrt{2}}
\end{pmatrix}
\begin{pmatrix}
0 &\nu \\
\nu & 0
\end{pmatrix}
\begin{pmatrix}
\frac{1}{\sqrt{2}} & \frac{ \mathrm{i}}{\sqrt{2}} \\
\frac{1}{\sqrt{2}} & -\frac{ \mathrm{i}}{\sqrt{2}}
\end{pmatrix}\notag\\
&= \begin{pmatrix}
\nu & 0 \\
0 & \nu
\end{pmatrix} = B_D.
\end{align}

This is the desired parallel single-mode squeezing operation~\cite{Ferraro2005}. Combining this with the CZ Bloch-Messiah reduction (Eqn.~\ref{eqn:BlochMessiah}), the CZ operation can therefore be implemented by $A = U' A_{TMS} V'^\dag, \, B = U' B_{TMS} V'^T$ where
\begin{align}
&U' = \begin{pmatrix}
\cos \phi' & -\sin \phi' \\
\mathrm{i} \sin \phi' & \mathrm{i} \cos \phi'
\end{pmatrix},\quad
V'=
\begin{pmatrix}
\cos \phi' & \sin \phi' \\
-\mathrm{i} \sin \phi' & \mathrm{i} \cos \phi'
\end{pmatrix},\notag\\
&A_{TMS} =
\begin{pmatrix}
\frac{\sqrt{5}}{2} & 0 \\
0 & \frac{\sqrt{5}}{2}
\end{pmatrix},\quad
B_{TMS} =
\begin{pmatrix}
0 & \frac{1}{2}\\
\frac{1}{2} & 0
\end{pmatrix}\label{eqn:CZ}
\end{align}
and $\phi' = \frac{\pi}{4} - \phi \approx 13.28^\circ$.

\begin{figure}[htbp]
\begin{center}
\includegraphics[width=7.5cm]{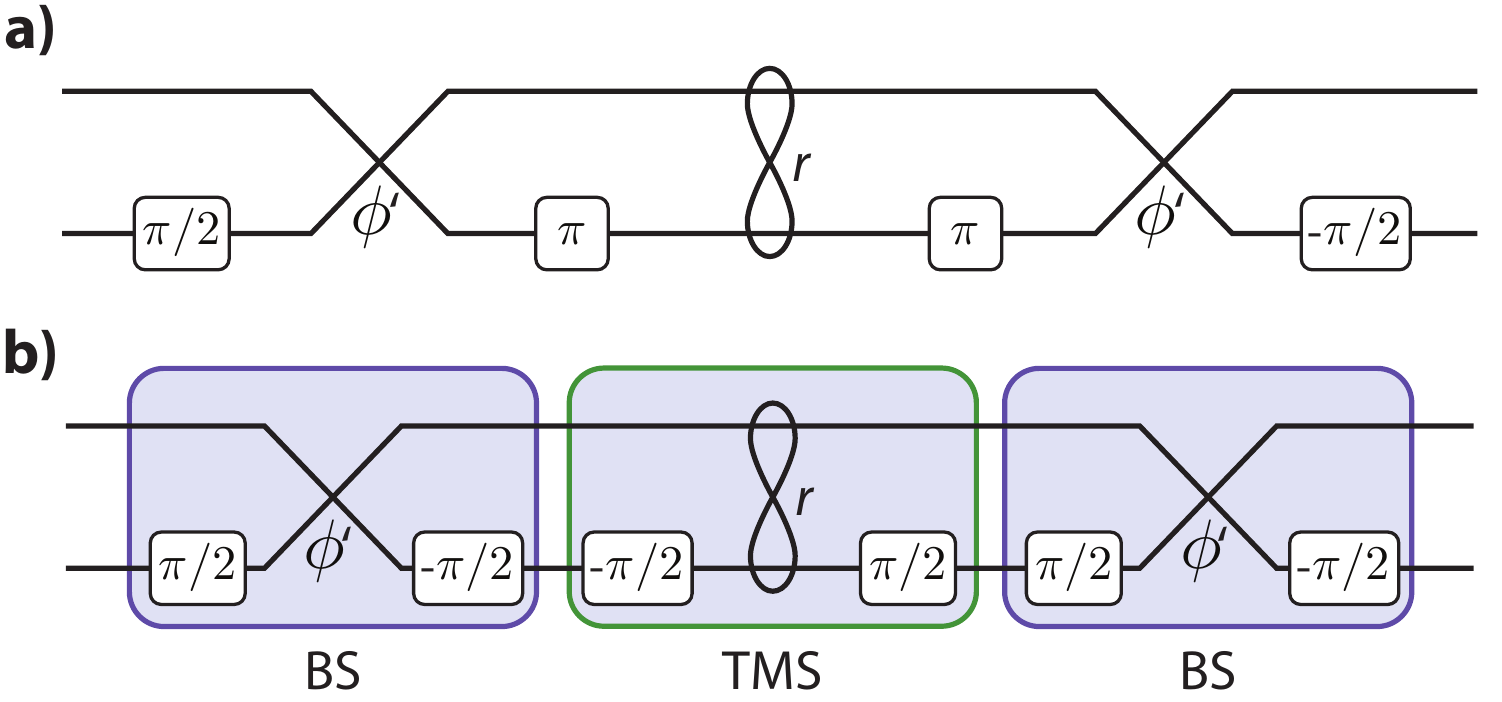}
\caption{a) The sequence of two-mode transformations necessary to implement a CZ operation, with $r = \sinh^{-1}(1/2)$, $\phi' = \frac{\pi}{4} - \frac{1}{2} \sin^{-1}(2/\sqrt{5})$. b) Three Raman memory operations are sufficient to implement this operation.}
\label{fig:CZdecomp}
\end{center}
\end{figure}

This operation is equivalent to the sequence of two-mode transformations shown in Fig~\ref{fig:CZdecomp}.

\FloatBarrier

\section{Memory requirements for cluster state generation}

The mode coupling BS transformations in the CZ operation (Fig~\ref{fig:CZdecomp}b) require a storage efficiency of $5.3\%$, and the TMS transformation requires 4.2 dB of squeezing. This is compatible with current state-of-the-art Raman memories, which have achieved storage efficiencies of over 80\%~\cite{Hosseini2011a}, and are estimated to be able to achieve a squeezing of 9.6 dB~\cite{Datta2012} (assuming that the modes for each transformation are similarly detuned from resonance).

This current state-of-the-art squeezing is below the levels needed for fault-tolerant quantum computing in the Menicucci scheme~\cite{Menicucci2014} (approximately 17-20 dB). However, in principle, an increase of 80\% in the mode coupling strength would be sufficient to provide this. With this level of increase, it would also be theoretically possible to transfer modes between electromagnetic field modes and the memory mode with unit efficiency.

In reality, all physical experimental implementations are imperfect at some level. For any real quantum memory, it likely that it is not the coupling strength, but rather the imperfect overlap between the input and output field modes, and other losses associated with field propagation, that are likely to ultimately limit the storage efficiencies achievable. It should be noted that these imperfections are obstacles in all quantum linear optical schemes, and as such are certainly not confined to memories as optical elements. However, in common with other such schemes, our proposal is sensitive to these deviations from ideal operation. Although a full treatment of the effects of these imperfections is beyond the scope of this work, in the following section we present calculations as an initial exploration of their expected impact.

\subsection{Fidelity of cluster state generation with imperfect memories}

We model imperfect two-qumode cluster state generation by applying a non-ideal read-out BS unitary (with a deviation from ideal operation parameterised by $\delta \eta$)
\begin{align}
\begin{pmatrix}
\sqrt{1 - \delta \eta}  & \sqrt{\delta \eta} \\
\sqrt{\delta \eta} & \sqrt{1 - \delta \eta}
\end{pmatrix} \label{eqn:imperfectBS}
\end{align}
between the memory mode of the initial generated TMS state (main text Fig.~3b) and a vacuum electromagnetic field mode.  We use the results of~\cite{Marian2012} to calculate the fidelity between this state and the ideal state, which we find to be
\begin{align}
4/\sqrt{\left(\sqrt{1-\delta \eta }-\left(\sqrt{1-\delta \eta }-1\right) \cosh (2 r)+1\right)^4}
\end{align}
where $r = -\frac{1}{2}\text{log}\left(10^{-\frac{dB}{10}}\right)$ and dB is the squeezing in decibels.

The fidelity gives a useful indication of the expected performance of this state for quantum computing as it can be shown to be equal to the minimum overlap of the probability distributions for the outcomes of any general measurement on the respective states~\cite{Barnum1996}. It therefore gives a rough estimate of the probability that a measurement of the state will give an erroneous result.

\begin{figure}[htbp]
\begin{center}
\includegraphics[width=8.5cm]{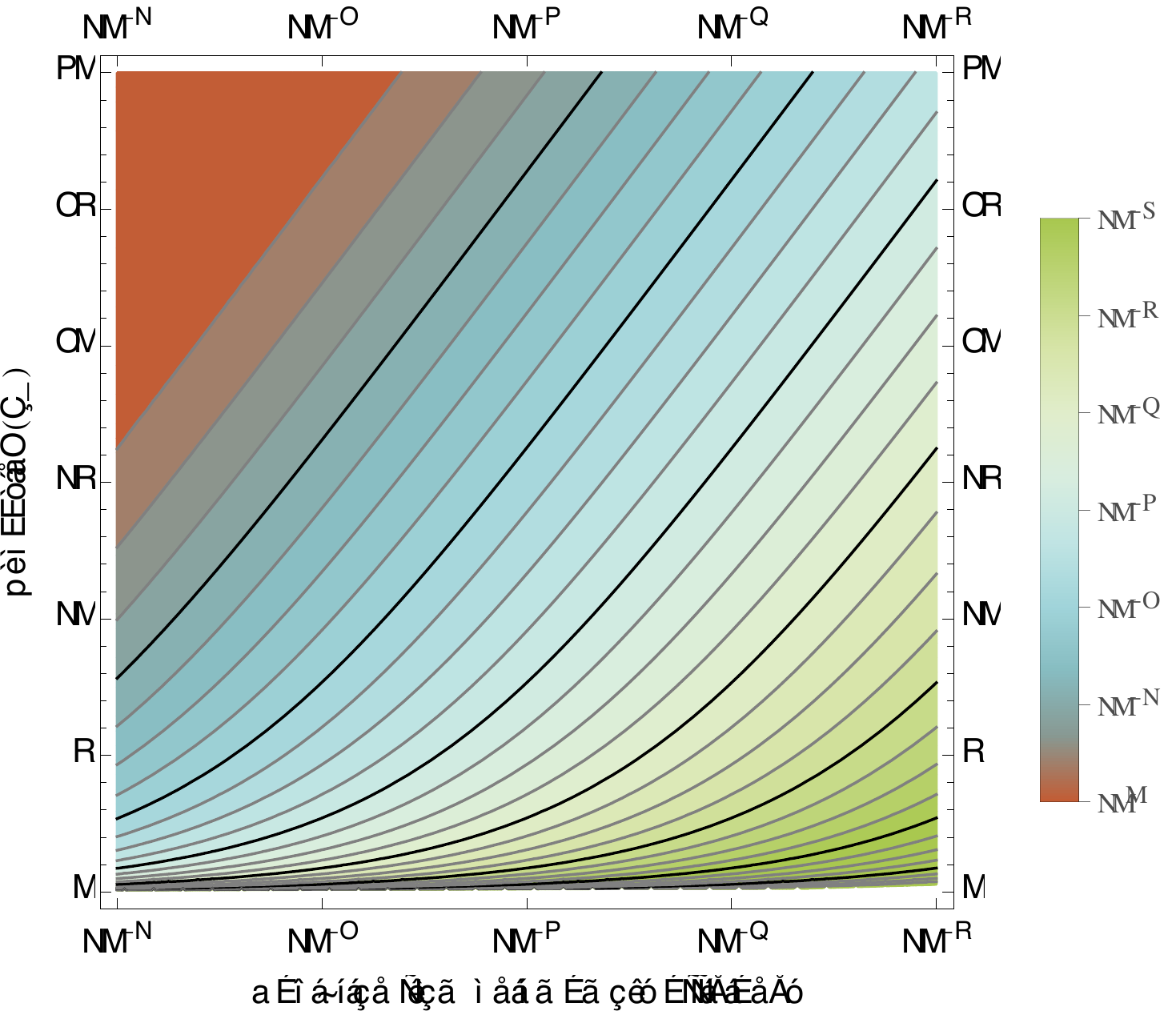}
\caption{The fidelity between the state generated by our two-qumode cluster state protocol and an ideal TMS state with the same level of squeezing, as a function of the squeezing desired and the deviation from unit memory efficiency for the read-out operation. Contours are plotted on a logarithmic scale, with solid black lines at each order of magnitude.}
\label{fig:UhlmannFidelity2ModeGen}
\end{center}
\end{figure}

Fig.~\ref{fig:UhlmannFidelity2ModeGen} shows a plot of this fidelity as a function of the desired squeezing level and the deviation from unit efficiency $\delta \eta$ of the read-out operation. As would be expected, the fidelity is increasingly sensitive to $\delta \eta$ as the squeezing increases. As an example, we consider a 17.4 dB squeezed cluster-state resource, sufficient in principle for fault-tolerant cluster-state computing if used with an error code able to tolerate gate error probabilities of $10^{-3}$~\cite{Menicucci2014}. In order to produce a two-qumode cluster with a corresponding $10^{-3}$ infidelity to the ideal state, it is necessary to be able to read out the state with $1 - 7 \times 10^{-5}$ efficiency. 

This is beyond the current state of the art. However, it is likely that this situation will be improved upon in the near future, both in terms of the requirements for fault tolerance, and in terms of the efficiencies that can be achieved. Specifically considering the case of fault tolerance, it should be noted that Menicucci~\cite{Menicucci2014} considers in detail only one specific continuous-variable scheme for encoding qubits~\cite{Gottesman2001a}. Using this scheme, he estimates the errors induced by finite squeezing and compares these with the error thresholds of contemporary qubit error correction schemes, in order to show that fault tolerance is in principle possible. However, this initial treatment explores only one small part of the parameter space of possible encodings, computational approaches and fault tolerance methods, suggesting that improvements may be possible. This is an important problem to the continuous-variable community and one of active research~\cite{Ferrini2014, Kieling2007}.

\subsection{Fidelity of outputs from an imperfect CZ operation}

\begin{figure}[htbp]
\begin{center}
\includegraphics[width=7.5cm]{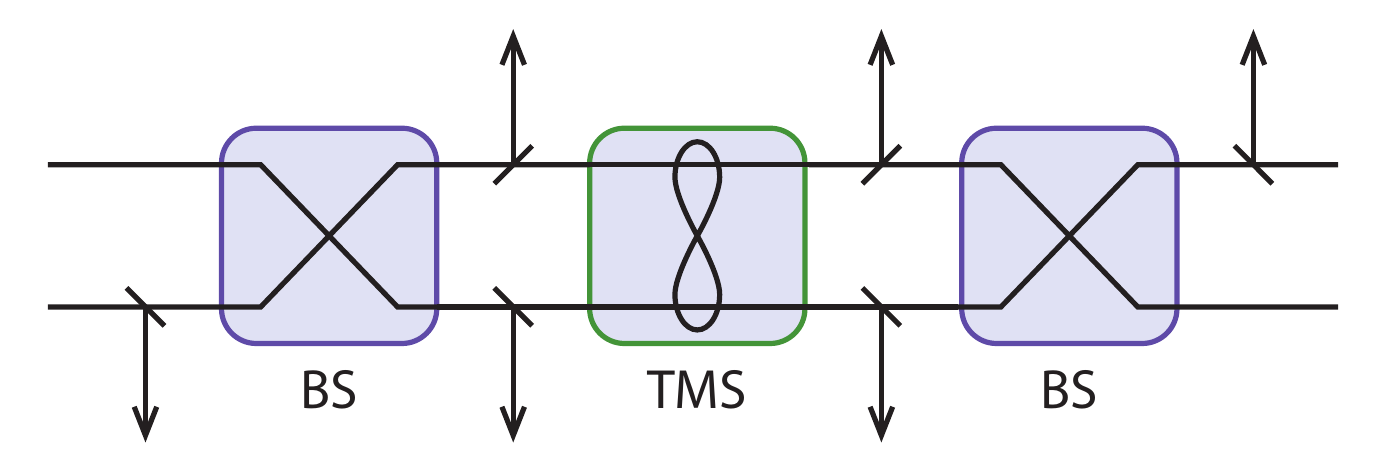}
\caption{Imperfect operation of the CZ operation protocol depicted in  Fig~4 of the main text is modelled by inserting loss modes (black arrows) at the positions depicted. These loss modes are coupled to the field modes by the BS unitary given in Eqn~\ref{eqn:imperfectBS}.}
\label{fig:CZGenCircuit}
\end{center}
\end{figure}

We additionally calculated the fidelity between the state output by an imperfect CZ operation used to entangle two uncorrelated single mode squeezed states and the equivalent ideal state. This is not the operation that will be carried out in our protocol, as that consists of linking two-qumode cluster state units. However, the additional complexity encountered in dealing with at least four field modes made this full calculation outside of the scope of this work. In our calculation, it is assumed that each of the BS operations that are used to transfer a qumode between a field mode and the spin wave mode in the CZ protocol (Fig~4 of the main text) deviates from perfect operation as defined in Eqn~\ref{eqn:imperfectBS}. This is equivalent to the set of transformations depicted in Fig~\ref{fig:CZGenCircuit}.

\begin{widetext}
For this CZ operation, we find a fidelity between the ideal and imperfect states of
\begin{align}
&4/\bigg(\Big(-(-2 (1-\delta \eta )^{3/2}+2 (1-\delta \eta )^5+(1-\delta \eta )^3+(1-\delta \eta )^2+2 \sqrt{1-\delta \eta }-1) (1-\delta \eta )-(((1-\delta \eta )^2\notag\\
&-6(1-\delta \eta )+2 \sqrt{1-\delta \eta }-2) (1-\delta \eta )+1) (1-\delta \eta ) \sinh (2 r)+(-2 (1-\delta \eta )^{5/2}+2 (1-\delta \eta )^6\notag\\
&+(1-\delta \eta )^4-\delta \eta +3) \cosh (2 r)+3 \Big) \Big(-2 (1-\delta \eta )^{3/2}+2 (1-\delta \eta )^{7/2}-2 (1-\delta \eta )^6-(1-\delta \eta )^5\notag\\
&-(1-\delta \eta )^3+(1-\delta \eta )^2+(((1-\delta \eta )^2 -2 (1-\delta \eta )+2 \sqrt{1-\delta \eta }-6) (1-\delta \eta )+1) (1-\delta \eta )^2 \sinh (2 r)\notag\\
&+(-2 (1-\delta \eta )^{7/2}+2 (1-\delta \eta )^6+(1-\delta \eta )^5+(1-\delta \eta )^2+2) \cosh (2 r)+3) \bigg)^{1/2}
\end{align}
\end{widetext}

\begin{figure}[htbp]
\begin{center}
\includegraphics[width=8.5cm]{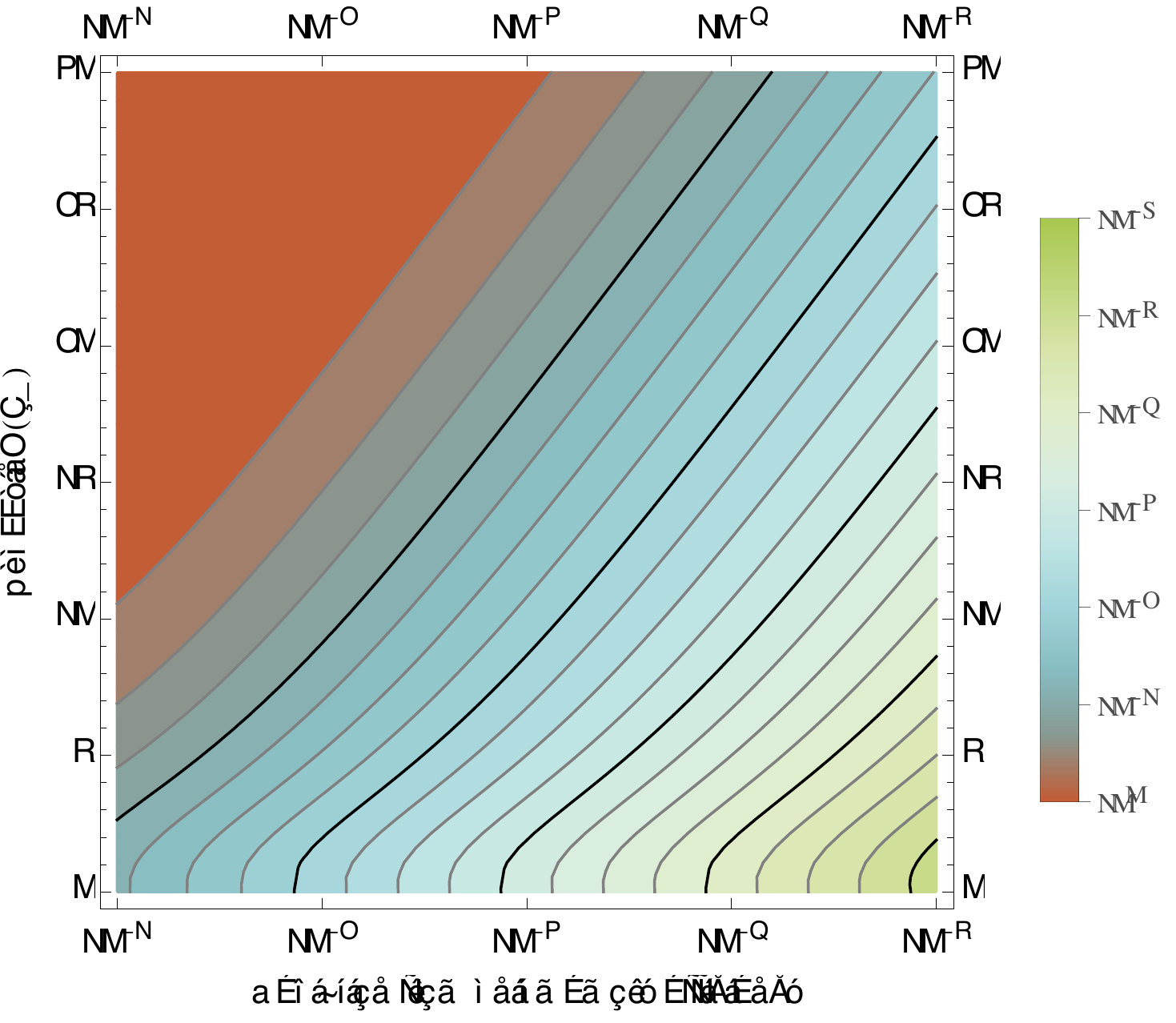}
\caption{The fidelity between the state generated by our two-qumode cluster state protocol and an ideal TMS state with the same level of squeezing as a function of the squeezing desired and the deviation from unit memory efficiency for the read-out operation. Contours are plotted on a logarithmic scale, with solid black lines at each order of magnitude.}
\label{fig:UhlmannFidelityCZ}
\end{center}
\end{figure}

Fig.~\ref{fig:UhlmannFidelityCZ} shows a plot of this fidelity as a function of the desired squeezing level and the deviation from unit efficiency $\delta \eta$ of the BS operations. We consider again a 17.4 dB squeezed cluster-state resource as an example. In order to entangle two qumodes with a corresponding $10^{-3}$ infidelity to the ideal state, we require BS operations with an efficiency of $1 - 10^{-5}$.

As can be seen from these calculations, the requirements for high fidelity operation at these squeezing levels are stringent. This presents a strong technical challenge for all experimental implementations of continuous-variable cluster state quantum computing schemes.

\FloatBarrier

\end{document}

%% file: header.tex
\usepackage{amsmath} 
\usepackage{amsthm} 
\usepackage{amssymb}	
\usepackage{graphicx} 

\newcommand{\ket}[1]{\left| #1 \right>} 
\let\baraccent=\= 
\renewcommand{\=}[1]{\stackrel{#1}{=}} 

\theoremstyle{definition}

\theoremstyle{remark}
